\documentclass[10pt,
showpacs,
preprintnumbers,amsmath,amssymb,
aps,prd,nofootinbib,eqsecnum,a4paper]{revtex4}
\usepackage{epsfig}
\usepackage{graphicx,epsf}
\usepackage{color}
\usepackage{bm}
\usepackage{psfrag}
\usepackage[symbol]{footmisc}
\usepackage{tensor}
\usepackage{hyperref}
\def\be{\begin{equation}}
\def\ee{\end{equation}}
\def\bea{\begin{eqnarray}}
\def\eea{\end{eqnarray}}
\def\bi{\begin{itemize}}
\def\ei{\end{itemize}}
\def\p{\partial}

\def\H{{\cal H}}

\def\cs2{c_{\rm{s}}^2}
\newcommand\eq[1]{Eq.~(\ref{#1})}
\newcommand\eqs[1]{Eqs.~(\ref{#1})}
\def\dd{\text{d}}

\def\dVf{\delta^{(1)} V}
\def\dVs{\delta^{(2)} V}

\def\das{\delta^{(2)} d_{A}}
\def\E{{\cal E}}

\def\dnuf{{\delta^{(1)} \nu}}

\def\dnus{{\delta^{(2)} \nu}}

\def\ff{{\Phi_{1}}}
\def\pf{{\Psi_{1}}}
\def\fs{{\Phi_{2}}}
\def\ps{{\Psi_{2}}}
\def\beg{\begin{enumerate}}
\def\en{\end{enumerate}}

\def\M0{{\cal M}_0}

\def\H{\mathcal{H}}

\begin{document}
%%%%%%%%%%%%%%%%%%%%%%

\title{Galaxy number counts at second order in perturbation theory: a
  leading-order term comparison}

\author{Jorge L.~Fuentes$^{1}$ \footnote[3]{\href{mailto:j.fuentesvenegas@qmul.ac.uk}{j.fuentesvenegas@qmul.ac.uk}} }
\author{Juan Carlos~Hidalgo$^{2}$ \footnote[2]{\href{mailto:hidalgo@icf.unam.mx}{hidalgo@icf.unam.mx}}}
\author{Karim A.~Malik$^{1}$ \footnote[1]{\href{mailto:k.malik@qmul.ac.uk}{k.malik@qmul.ac.uk}}}
\affiliation{
$^1$Astronomy Unit, School of Physics and Astronomy, Queen Mary University of London, Mile End Road, London, E1 4NS, United Kingdom\\
$^2$Instituto de Ciencias F\'{i}sicas, Universidad Nacional Aut\'{o}noma de M\'{e}xico,\\Av. Universidad S/N. Cuernavaca, Morelos, 62251, M\'{e}xico}
\date{\today}

\begin{abstract}
  The galaxy number density is a key quantity to compare theoretical
  predictions to the observational data from current and future Large
  Scale Structure surveys. The precision demanded by these Stage IV
  surveys requires the use of second order cosmological perturbation
  theory. Based on the independent calculation published previously,
  we present the result of the comparison with the results of three
  other groups at leading order. 
  Overall we find that the differences between the different approaches lie mostly on the definition of certain quantities, where the ambiguity of signs results in the addition of extra terms at second order in perturbation theory.
  
\end{abstract}

\pacs{98.80.Cq \hfill  arXiv:2012.15326}

\maketitle

%%%%%%%%%%%%%%%%%%%%%%%%%%%%%%%%%%%%%
\section{Introduction}
%%%%%%%%%%%%%%%%%%%%%%%%%%%%%%%%%%%%%

Within the next couple of years next generation Large Scale Structure
(LSS) surveys such as Euclid \cite{euclid}, MeerKAT
\cite{santos2017meerklass}, SKA \cite{maartens2015cosmology}, the
VRO/LSST \cite{lsst2009}, DESI \cite{desi}, J-PAS \cite{jpas} and WFIRST \cite{WFIRST} will begin to take
data of unprecedented quality and quantity. It is therefore high time
for theoretical cosmologists to match the precision of their
predictions to that of the forthcoming Stage IV experiments.
%
%BOSS \cite{BOSS}, eBOSS
%\cite{eBOSS}, Euclid \cite{euclid}, MeerKAT \cite{santos2017meerklass},
%SKA \cite{maartens2015cosmology}, Rubin Obervatory/LSST
%\cite{lsst2009}, and WFIRST
%\cite{WFIRST} 
%
One of the key quantities in this context is the galaxy number
density, which allows us to compare theoretical predictions to the
observational data provided by the experiments. However, the
calculation of this quantity is rather involved, in particular if we
take the calculation beyond first order in cosmological perturbation
theory. Going beyond linear order is necessary to capture all the
subtle effects affecting the number density and to achieve the
precision required by the latest observational data.

Recently three different derivations of the second order number density
were published \cite{durrer2, cc1,cc2, yoo2}.
Unfortunately it quite difficult to compare the results of the three
derivations for the galaxy number density. First of all, the final
expressions for the galaxy number density occupy several pages. Then,
the notation and the break-up into separate terms is very different
throughout all the derivations. An additional difficulty is that terms can
be converted into each other by integration by parts, in a non-trivial
way, making the comparison of even partial results tricky at best.

Because of these difficulties, there is a general question of whether
or not all these derivations coincide. A first comparison of the
different derivations has been done in Ref.~\cite{nielsen}, where the authors
concentrate on the terms which dominate on sub-horizon scales,
considering only the terms of the order of $\left( k/\H
\right)^{4}\Psi^{2}$ and neglect smaller contributions to the second
order number count. Here, $k$ is the comoving wave number, $\H$ the
conformal Hubble parameter and $\Psi$ a scalar metric perturbation. In
Ref.~\cite{nielsen} the authors use a straight-forward derivation to
reproduce the result of Ref.~\cite{durrer2}, finding disagreements
concerning lensing terms and a double counting of volume distortion
effects in Refs.~\cite{cc1} and \cite{yoo2}, respectively.

Given the importance of the number density for current and future LSS
surveys, we decided to have an independent calculation of this
quantity up to second order in the perturbations and published the
results in Ref.~\cite{FMH}. In this paper, we use the derivation of
the second order number count given in Ref.~\cite{FMH} and follow a
similar approach as the one made in Ref.~\cite{nielsen} comparing
leading terms in the aforementioned literature, however not
constraining our comparison to sub-horizon scales

{\it{Notation.}} Greek indices $(\mu,\nu,\dots)$ denote space-time components, while Latin indices (i,j,\dots) stand for spatial components in a perturbed Friedmann-Lemaitre-Robertson-Walker
(FLRW) spacetime. A dash represents
derivative with respect to the conformal time $X' \equiv \frac{\p X}{\p\eta} $, while the derivative with respect to the affine parameter is  
\begin{equation}
\label{eq:der-affine}
\frac{\dd X}{\dd \lambda} = X' + n^{i}X_{,i}\,,
\end{equation}
where $n^{i}$ is the direction of observation (line of sight). For a scalar function $X$,
this implies, %
\begin{equation}
\label{eq:double-der-affine}
n^{i}n^{j}X_{,ij}= n^{i}n^{j}\p_{i}\p_{j}X
= \frac{\dd^{2} X}{\dd \lambda^{2}}-2 \frac{\dd X'}{\dd \lambda} +X''\,,
\end{equation}
where $\nabla_{i}X$ is the spatial part of the covariant derivative, and for the scalar $\nabla_{i}X = \partial_{i}X \equiv X_{,i}$. 

 Finally, when evaluating quantities at the source (s) and observer (o) points, we use the notation
\be
\label{eq:not1}
\left(X\right)^{s}_{o} = X\big|^{s}_{o} = X_{s}-X_{o} = X(\lambda_{s})-X(\lambda_{o})\,.
\ee
%
%

%%%%%%%%%%%%%%%%%%%%%%%%%%%%%%%%%%%%%
\section{Basic definitions for the galaxy number density}
\label{sec:whatwemeasure}
%%%%%%%%%%%%%%%%%%%%%%%%%%%%%%%%%%%%%

To exercise our notation and for convenience of use we define in this section
the basic quantities required to calculate the galaxy number density
up to second order in cosmological perturbation theory. We follow very
closely our previous paper, Ref.~\cite{FMH}. However, we do not
reproduce the rather lengthy calculations presented there, and focus
here on the results. For a more pedagogic treatment of the galaxy
number density calculation see Ref.~\cite{jorge_thesis}.

\subsection{Metric perturbations}

The perturbed FLRW spacetime is described in the longitudinal gauge by \cite{malik}
\be
\label{eq:metric}
\dd s^{2} = a^{2}\left[ -\left(1+2\ff+\fs \right) \dd \eta^{2}
  + \left( 1 - 2 \pf - \ps \right) \delta_{ij} \dd x^{i}\dd x^{j}\right]\,,
\ee
where $a=a(\eta)$ is the scale factor, and $\eta$ is the conformal time;
$\delta_{ij}$ is the flat spatial metric, and we have neglected
the vector and tensor modes, we also allow for first and second order
anisotropic stresses. From now on we consider perturbations around a
FLRW metric up to second-order.

\subsection{Matter velocity field and peculiar velocities}

The components of the four-velocity $u^{\mu}=\dd x^{\mu}/\dd \eta$, up to
second order, for the perturbed metric are 
\begin{align}
\label{eq:pertu0}
u_{0} &= -a \left[ 1 + \ff + \frac{1}{2} \fs- \frac{1}{2}\ff^{2}+\frac{1}{2}v_{1 k}v_{1}^{k} \right], \\
\label{eq:pertui}
u_{i} &= a \left[ v_{1 i} + \frac{1}{2} v_{2 i}  - 2 \pf v_{1 k}\right], \\
u^{0} &= a^{-1} \left[1-\ff - \frac{1}{2}\fs+\frac{3}{2} \ff^{2}+\frac{1}{2}v_{1 k}v_{1}^{k} \right],\\
u^{i} &= a^{-1} \left[ v_{1}^{i} + \frac{1}{2} v_{2}^{i} \right]\,,
\end{align}
where $v_{i} = \partial_{i} \text{v}$, and with $\text{v}$ the velocity potential.

\subsection{Photon wavevector}

In a redshift survey galaxy positions are identified through the detection of photons emitted by a source, denoted by $s$,  and detected by the observer, labelled $o$. In a general spacetime, we consider a lightray with tangent vector $k^{\mu}$ and affine parameter $\lambda$, that parametrises the trajectory followed by the lightray. The source and the observer are represented by specific values of the affine parameter, $\lambda_{s}$ and $\lambda_{o}$, respectively. The components of the photon wavevector are
\be
\label{eq:kbg}
\bar{k}^{\mu} = \frac{\dd x^{\mu}}{\dd \lambda} = a^{-1}\Big[1,n^{i}\Big]\,,
\ee
where the overbar denotes background quantities. The direction of observation is given by the vector $n^{i}$ which
points from the observer to the source\footnote{Some authors define $n^{i}$ with the
  opposite sign. See, for example, Refs.~\cite{durrer1,cc1,cc2}.}, and obeys the normalisation condition: $n^{i}n_{i} = 1$.

The tangent vector is \textit{null} 
\be
\label{eq:null}
k_{\mu}k^{\mu} = 0\,,
\ee
and \textit{geodesic} 
\be
\label{eq:geodesic}
k^{\nu}\nabla_{\nu}k^{\mu} =0\,,
\ee
where $\nabla_{\nu}$ is the covariant derivative defined by the metric given in \eq{eq:metric}.
To make a distinction, we denote the perturbed wavevector as
\be
\delta^{(n)} k^{\mu}=a^{-1}\Big[\delta^{(n)} \nu, \delta^{(n)} n^{i}\Big]\,,
\ee
where $\delta^{(n)}$ gives the $n\text{-th}$ order perturbation, and where 
the usual notation is followed for the temporal component, that
is $k^{0}\equiv\nu$ (see e.g.~Ref.~\cite{yoo4}).

The affine parameter of the geodesic equation is also related to the
comoving distance ($\chi$) by
\be
\label{eq:comovingd}
\chi = \lambda_{o}-\lambda_{s}\,,
\ee
which in terms of the redshift is
\be
\label{eq:comd}
\chi(z) = \int_{0}^{z} \frac{\dd \tilde{z}}{(1+\tilde{z})\H (\tilde{z})}\,.
\ee

%%%%%%%%%%%%%%%%%%%%%%%%%%%%%%%%%%%%%
\subsection{The observed redshift}
%%%%%%%%%%%%%%%%%%%%%%%%%%%%%%%%%%%%%

The photon energy measured by an observer with 4-velocity $u^{\mu}$ is 
\be
\label{eq:photon-energy}
\E = - g_{\mu \nu}u^{\mu}k^{\nu}\,.
\ee
This implies that the observed redshift of a source (e.g.~a galaxy) can be defined as
\be
\label{eq:redshift}
1+z = \frac{\E_{s}}{\E_{o}} \,.
\ee
\noindent This definition explicitly shows the nature of the  Doppler effect on  the observed redshift, a function of the velocity and the wavevector, i.e.~$z=z(k^{\mu},u^{\mu})$.

%%%%%%%%%%%%%%%%%%%%%%%%%%%%%%%%%%%%%
\subsection{Angular diameter distance}
%%%%%%%%%%%%%%%%%%%%%%%%%%%%%%%%%%%%%

A given bundle of lightrays leaving a source will invariantly expand and create a distance in between the lightrays that
conform it, this can be projected to an area in screen space,
the hypersurface perpendicular to the trajectories of the photons and the 4-velocity of
the observer. If we denote the area of a bundle in screen space with $\mathcal{A}$, then the
angular diameter distance $d_{A}$, and the null expansion $\theta$ are derived from this quantity as \cite{cc4},
\begin{equation}
\label{eq:daa}
\frac{1}{\sqrt{\mathcal{A}}}\frac{\dd \sqrt{\mathcal{A}}}{\dd \lambda}
= \frac{\dd \ln d_{A}}{\dd \lambda} = \frac{1}{2} \theta\,.
\end{equation}
 From this we can compute how the area of the bundle changes along
the geodesic trajectory followed by the photons from the source
up to the observer.

%%%%%%%%%%%%%%%%%%%%%%%%%%%%%%%%%%%%%
\subsection{Physical Volume}
%%%%%%%%%%%%%%%%%%%%%%%%%%%%%%%%%%%%%

Number counts account for the number of sources detected in a bundle of
rays, for a small affine parameter displacement $\lambda$ to
$\lambda+\dd\lambda$ at an event $P$. This defines the physical
distance
\be
\label{eq:dl}
\dd \ell = (k^{\mu}u_{\mu}) \dd \lambda\,,
\ee 
in the rest frame of a comoving galaxy at said point in space $P$, and is positive if $k^{\mu}$ is a tangent vector to the past directed null geodesics,  so that $k^{\mu}u_{\mu}>0$.

On the other hand, the cross-sectional area of the bundle is 
\be
\label{eq:cross-a}
\dd \mathcal{A} = d_{A}^{2}(\lambda)\dd \Omega\,,
\ee 
if the geodesics subtend a solid angle $\dd \Omega$ at the observer.

From \eqs{eq:dl} and \eqref{eq:cross-a} the corresponding volume
element at a point $P$ in space is (see e.g.~Ref.~\cite{ellis})
\begin{equation}
\label{eq:dldo}
\dd V = \dd \ell \dd \mathcal{A}
= (k^{\mu}u_{\mu})d_{A}^{2}(\lambda) \dd \lambda \dd \Omega
= -\E d^{2}_{A}(\lambda) \dd \lambda \dd \Omega\,.
\end{equation}
These covariant definitions are used in a second order expansion of the Cosmological perturbation theory in the
next section.

%%%%%%%%%%%%%%%%%%%%%%%%%%%%%%%%%%%%%
\section{Leading-order terms at second order}
\label{sec:dominant-terms}
%%%%%%%%%%%%%%%%%%%%%%%%%%%%%%%%%%%%%

To determine the leading terms of the second order number counts, we
first need to analyse the result given in Eq.~(5.17) from
Ref.~\cite{FMH}. From Eqs.~(5.8) and (5.9) therein, we can see that
the general expression for the dominating terms is of the form
\be
\label{eq:dleading}
\Delta_{\text{Leading}} = \left( 1 + \delta \right)\left( 1 + \delta V \right) \simeq \left(1+\delta\right)\left(1+\text{RSD}\right)\left(1+\kappa\right)\,,
\ee
where $\delta$ is the matter overdensity, $\delta V$ is the perturbed
volume, and we further split the volume perturbation into its dominant
components; that is, the lensing contribution ($\kappa$), and redshift
space distortions (RSD).

Perturbing Eq.~\eqref{eq:dleading} up to second order in perturbation
theory, we find that the nonlinear contribution to the leading terms
is given by
\begin{align}
\label{eq:leadingdg}
\Delta^{(2)}_{\text{Leading}} &= \delta_{g}^{(2)} + \delta_{g}^{(1)}
\dVf + \dVs, \\ &= \delta_{g}^{(2)} + \left[\text{RSD}\right]^{(2)}+
\kappa^{(2)} + \delta_{g}^{(1)}\left[ \text{RSD}\right]^{(1)} +
\delta_{g}^{(1)}\kappa^{(1)} + \left[ \text{RSD}
  \right]^{(1)}\kappa^{(1)}\,. \notag
\end{align}

We now present the leading terms of the calculation to second order
given in Eq.~(5.17) from Ref.~\cite{FMH}, which is
\begin{align}
&\Delta^{(2)}_{\text{Leading}} \simeq \delta_{g}^{(2)} -\frac{1}{2\H}\frac{\dd\left( v_{2i}n^{i}\right)}{\dd\varsigma} + \frac{1}{2\H}\frac{\dd\left(v_{1k}v_{1}^{k}\right)}{\dd\varsigma}-\left(v_{1i}n^{i}\right)\frac{\dd\left(v_{1i}n^{i}\right)}{\dd\varsigma}\\
& +\frac{\dd\left(v_{1i}n^{i}\right)}{\dd\varsigma}\int_{0}^{\chi_{s}}\dd\chi\left( \ff'+\pf' \right)-\frac{\H''}{\H^{3}}\frac{\dd\left(v_{1i}n^{i}\right)}{\dd\varsigma}\int_{0}^{\chi_{s}}\dd\chi\left( \ff'+\pf' \right) \notag\\
& +\frac{2}{\H}\frac{\dd\left(v_{1i}n^{i}\right)}{\dd\varsigma} \int_{0}^{\chi_{s}}\dd\chi\left( \ff'+\pf' \right) +\frac{\delta_{g}^{(1)}}{\H}\frac{\dd \left(v_{1i}n^{i}\right)}{\dd\varsigma}\notag\\
&+ \frac{1}{\H}\frac{\dd\left(v_{1i}n^{i}\right)}{\dd\varsigma}\frac{1}{\chi}\int_{0}^{\chi_{s}}\dd\chi\left[ \nabla^{2}\left( \ff+\ff \right) + n^{i}n^{j}\left( \ff+\pf \right)_{,ij}+\frac{2}{\chi}\frac{\dd\dnuf}{\dd\varsigma}\right] \notag\\
& +\frac{\delta_{g}^{(1)}}{\chi}\int_{0}^{\chi_{s}}\dd\chi\left[ \nabla^{2}\left( \ff+\pf \right) + n^{i}n^{j}\left( \ff+\pf \right)_{,ij}+\frac{2}{\chi}\frac{\dd\dnuf}{\dd\varsigma}\right] \notag\\
& +\frac{2}{\chi} \int_{0}^{\chi_{s}}\dd\chi\left[ \nabla^{2}\left( \ps+\fs \right) + n^{i}n^{j}\left( \fs+\ps \right)_{,ij}+\frac{2}{\chi}\frac{\dd\dnus}{\dd\varsigma}\right]\,. \notag 
\end{align}

These terms originate from density fluctuations\footnote{For a work discussing the second-order density perturbation in the case of interacting vacuum cosmologies see Ref.~\cite{Borges:2017jvi}.},
\begin{align}
\label{eq:delta-2}
&\Delta^{(2)}_{\text{Leading--}\delta} \simeq \delta_{g}^{(2)}\,,
\end{align}
redshift space distortions 
\begin{align}
\label{eq:RSD-2}
&\Delta^{(2)}_{\text{Leading--RSD}} \simeq -\frac{1}{2\H}\frac{\dd\left( v_{2i}n^{i}\right)}{\dd\varsigma} + \frac{1}{2\H}\frac{\dd\left(v_{1k}v_{1}^{k}\right)}{\dd\varsigma}-\left(v_{1i}n^{i}\right)\frac{\dd\left(v_{1i}n^{i}\right)}{\dd\varsigma}\\
& \qquad\qquad\qquad\qquad +\left[1-\frac{\H''}{\H^{3}}+\frac{2}{\H}\right]\frac{\dd\left(v_{1i}n^{i}\right)}{\dd\varsigma} \int_{0}^{\chi_{s}}\dd\chi\left( \ff'+\pf' \right)\,, \notag
\end{align}
lensing terms
\begin{align}
&\Delta^{(2)}_{\text{Leading--}\kappa} \simeq \frac{2}{\chi} \int_{0}^{\chi_{s}}\dd\chi\left[ \nabla^{2}\left( \ps+\fs \right) + n^{i}n^{j}\left( \fs+\ps \right)_{,ij}+\frac{2}{\chi}\frac{\dd\dnus}{\dd\varsigma}\right]\,,  
\label{eq:k-2}
\end{align}
and cross terms
\begin{align}
\label{eq:delta-RSD}
&\Delta^{(2)}_{\text{Leading--}\delta\times \text{RSD}} \simeq \frac{\delta_{g}^{(1)}}{\H}\frac{\dd \left(v_{1i}n^{i}\right)}{\dd\varsigma}, \\
&\Delta^{(2)}_{\text{Leading--}\delta\times\kappa} \simeq \frac{\delta_{g}^{(1)}}{\chi}\int_{0}^{\chi_{s}}\dd\chi\left[ \nabla^{2}\left( \ff+\pf \right) + n^{i}n^{j}\left( \ff+\pf \right)_{,ij}+\frac{2}{\chi}\frac{\dd\dnuf}{\dd\varsigma}\right], \\
&\Delta^{(2)}_{\text{Leading--RSD}\times \kappa} \simeq \frac{1}{\H}\frac{\dd\left(v_{1i}n^{i}\right)}{\dd\varsigma}\frac{1}{\chi}\int_{0}^{\chi_{s}}\dd\chi\left[ \nabla^{2}\left( \ff+\ff \right) + n^{i}n^{j}\left( \ff+\pf \right)_{,ij}+\frac{2}{\chi}\frac{\dd\dnuf}{\dd\varsigma}\right]\,.
\end{align}

%%%%%%%%%%%%%%%%%%%%%%%%%%%%%%%%%%%%
\section{Comparison of the leading order terms}
\label{sec:secondcomparison}
%%%%%%%%%%%%%%%%%%%%%%%%%%%%%%%%%%%%

%%%%%%%%%%%%%%%%%%%%%%%%%%%%%%%%%%%%%
\subsection{Comparison with Di Dio, et al.}
%%%%%%%%%%%%%%%%%%%%%%%%%%%%%%%%%%%%%

In the following, we identify and match terms from Ref.~\cite{durrer2}
to Eq.~(5.17) from Ref.~\cite{FMH}. We start by relating the notation
of one to the other, the main notational differences between our work
and Ref.~\cite{durrer2} are:
\begin{itemize}
\item Work in the geodesic light-cone gauge (GLC) to obtain their solution which is then expanded in a more conventional gauge, Poisson gauge, to second order.
\item Latin indices in the GLC take only the values 1, 2.
\item Maintain their integrals in terms of the conformal time $\eta$.
\item Separate the final result into an isotropic and anisotropic part.
\item Projected notation is used $v_{\parallel} = n^{i}v_{i}$.
\item There is a factor of $(-1)$ in the definition of the observation vector $n^{i}$.
\end{itemize}

To second order, the main result is given by Eq.~(4.41) in their paper,
\be
\Delta^{(2)} = \Sigma - \langle \Sigma \rangle\,,
\ee
where
\be
\Sigma = \Sigma_{\text{IS}} + \Sigma_{\text{AS}}\,,
\ee
and the isotropic, $\Sigma_{\text{IS}}$, and anisotropic,
$\Sigma_{\text{AS}}$, parts are given in Eqs.~(4.42) and (4.43) of
Ref.~\cite{durrer2}, respectively.

The translation of the majority of the terms is straightforward. $r$
is the comoving distance we call $\chi$ and $\partial_{\eta} =
\dd/\dd\eta = \dd/\dd\varsigma - n^{i}\partial_{i} =
-\dd/\dd\chi$. Thankfully, the authors of Ref.~\cite{durrer2} include
a \textit{leading} order second order contribution in their paper,
which is
\begin{align}
&{}^{(2)}\Delta^{\text{Di Dio}}_{\text{Leading}} \simeq \delta_{\rho}^{(2)} + \frac{1}{\H}\frac{\dd \left( v_{2i}n^{i}\right)}{\dd\varsigma} -\frac{1}{2 r} \int_{\eta_{s}}^{\eta_{o}}\dd\eta'\frac{\eta'-\eta_{s}}{\eta_{o}-\eta'} \Delta_{2}\left( \ps + \fs \right) \\
&+\frac{1}{\H}\left[ \left( v_{1i}n^{i} \right)\frac{\dd^{2}\left( v_{1i}n^{i} \right)}{\dd \varsigma^{2}} + \left( \frac{\dd\left( v_{1i}n^{i} \right)}{\dd\varsigma} \right)^{2} \right] - \frac{2}{\H}\frac{\dd\left( v_{1i}n^{i}\right)}{\dd\varsigma} \frac{1}{r}\int_{\eta_{s}}^{\eta_{o}}\dd\eta'\frac{\eta'-\eta_{s}}{\eta_{o}-\eta'}\Delta_{2}\psi^{I} \notag \\
& -\frac{2}{\H}\partial_{a}\left( \frac{\dd\left( v_{1i}n^{i} \right)}{\dd\varsigma} \right)\int_{\eta_{s}}^{\eta_{o}}\dd\eta' \gamma_{0}^{ab}\partial_{b}\int_{\eta_{s}}^{\eta_{s}}\dd\eta''\psi^{I} +2\left( \frac{1}{r}\int_{\eta_{s}}^{\eta_{o}}\dd\eta'\frac{\eta'-\eta_{s}}{\eta_{o}-\eta'}\Delta_{2}\psi^{I} \right)^{2} \notag\\
& +\frac{4}{r}\int_{\eta_{s}}^{\eta_{o}}\dd\eta'\frac{\eta'-\eta_{s}}{\eta_{o}-\eta'}\Bigg\{\partial_{b}\left[\Delta_{2}\psi^{I}\right]\int_{\eta_{s}}^{\eta_{o}}\dd\eta''\gamma_{0}^{ab}\partial_{a}\int_{\eta''}^{\eta_{o}}\dd\eta'''\psi^{I}  \notag \\
& +\Delta_{2}\left[ -\frac{1}{2}\gamma_{0}^{ab}\partial_{a}\left( \int_{\eta'}^{\eta_{o}}\dd\eta''\psi^{I} \right)\partial_{b}\left( \int_{\eta'}^{\eta_{o}}\dd\varsigma''\psi^{I} \right) \right]\Bigg\} \notag\\
& -4\int_{\eta_{s}}^{\eta_{o}}\dd\eta'\Big\{ \gamma_{0}^{ab}\partial_{b}\left( \int_{\eta'}^{\eta_{o}}\dd\eta''\psi^{I} \right)\frac{1}{\eta_{o}-\eta'}\int_{\eta'}^{\eta_{o}}\dd\eta''\frac{\eta''-\eta'}{\eta_{o}-\eta''}\partial_{a}\Delta_{2}\psi^{I} \Big\}\notag\\
& +\left[ \frac{1}{\H}\left( \frac{\dd\left( v_{1i}n^{i} \right)}{\dd\varsigma}\right) - \frac{2}{r}\int_{\eta_{s}}^{\eta_{o}}\dd\eta'\frac{\eta'-\eta_{s}}{\eta_{o}-\eta'}\Delta_{2}\psi^{I} \right]\delta_{\rho}^{(1)} +\frac{1}{\H}\left( v_{1i}n^{i} \right)\frac{\dd \delta_{\rho}^{(1)}}{\dd\varsigma} \notag \\
& -2\partial_{a}\delta_{\rho}^{(1)}\int_{\eta_{s}}^{\eta_{o}}\dd\eta'\gamma_{0}^{ab}\partial_{b}\int_{\eta'}^{\eta_{o}}\dd\eta\psi^{I}. \notag
\end{align}
After some integrations by parts to rewrite some terms, and rewriting what the angular Laplacian is in our notation, the only difference between our leading terms and the ones from Ref.~\cite{durrer2} comes from the oposite sign in the definition of the observation vector $n^{i}$, and a numerical factor between the four velocities, one half.
\begin{align}
\Delta^{(2)}_{\text{Leading}} - {}^{(2)}\Delta^{\text{Di Dio}}_{\text{Leading}} &\simeq \Big[\delta_{g}^{(2)} - \delta_{\rho}^{(2)}\Big] - \frac{1}{\H}\frac{\dd\left( v_{2i}n^{i}\right)}{\dd\varsigma}\left[ \frac{1}{2} + 1 \right], \\
&\simeq - \frac{1}{\H}\frac{\dd\left( v_{2i}n^{i}\right)}{\dd\varsigma}\left[ \frac{1}{2} + 1 \right]\,, \notag
\end{align}
where we can see explicitly that the difference comes from the
opposite sign in the definition of the observation vector $n^{i}$,
and a $(1/2)$-factor in the definition of our second order
velocity perturbation. 

%%%%%%%%%%%%%%%%%%%%%%%%%%%%%%%%%%%%%
\subsection{Comparison with Bertacca, et al.}
%%%%%%%%%%%%%%%%%%%%%%%%%%%%%%%%%%%%%

We now proceed to identify the terms of Refs.~\cite{cc1,cc2}, the main remarks about the notation here are:
\begin{itemize}
\item They work with cosmic rulers, then change to Poisson gauge.
\item There is a factor of $(-1)$ in the definition of the observation vector $n^{i}$.
\item Projected derivatives, meaning $\partial_{\parallel}=n^{i}\partial_{i}$ and $\partial_{\perp}^{i} = r^{-1}\partial^{i}$.
\end{itemize}

The leading terms in the expression for the number counts, would be,
\begin{align}
\label{eq:bert2}
&{}^{(2)}\Delta^{\text{Bertacca}}_{\text{Leading}} \simeq \delta_{g}^{(2)} - \frac{1}{\H} \frac{\dd^{2} \left( v_{2i}n^{i}\right)}{\dd\varsigma^{2}} - 2\kappa^{(2)} + 4\left[ \kappa^{(1)} \right]^{2} - 4\delta_{g}^{(1)}\kappa^{(1)} \\
&- 2 \frac{\delta_{g}^{(1)}}{\H} \frac{\dd^{2}\left( v_{1i}n^{i} \right)}{\dd\varsigma^{2}} + 4\frac{\kappa^{(1)}}{\H}\frac{\dd^{2}\left( v_{1i}n^{i} \right)}{\dd\varsigma^{2}} + \frac{2}{\H^{2}}\left[ \frac{\dd^{2}\left( v_{1i}n^{i} \right)}{\dd\varsigma^{2}} \right]^{2} + \frac{2}{\H^{2}}\left[\frac{\dd\left( v_{1i}n^{i} \right)}{\dd\varsigma}\right]\left[ \frac{\dd^{3}\left( v_{1i}n^{i} \right)}{\dd\varsigma^{3}}\right] \notag\\
& -\frac{2}{\H}\frac{\dd \delta_{g}^{(1)}}{\dd\chi}\Delta\ln a^{(1)} + \frac{2}{\chi}\left[ \left( \delta_{g}^{(1)}\right)_{,i} -\frac{1}{\H}\frac{\dd^{2}\left(v_{1k}n^{k}\right)_{,i}}{\dd\varsigma^{2}} \right]\int_{0}^{\chi_{s}}\dd\chi\left( {\ff_{,}}^{i} + {\pf_{,}}^{i} \right) \notag \\
& -2\left[ \left( \delta_{g}^{(1)}\right)_{,i} -\frac{1}{\H}\frac{\dd^{2}\left(v_{1k}n^{k}\right)_{,i}}{\dd\varsigma^{2}} \right] \int_{0}^{\chi_{s}}\dd\chi\frac{1}{\chi}\left( {\ff_{,}}^{i} + {\pf_{,}}^{i} \right)\notag\\
&-4\left( \int_{0}^{\chi_{s}}\dd\tilde{\chi}\frac{\tilde{\chi}}{\chi}\left( \bar{\chi}-\tilde{\chi} \right)\mathcal{P}^{n}_{i}\mathcal{P}^{jm}\partial_{m}\partial_{n}\ff \right)\left( \int_{0}^{\chi_{s}}\dd\tilde{\chi}\frac{\tilde{\chi}}{\chi}\left( \bar{\chi}-\tilde{\chi} \right)\mathcal{P}^{p}_{i}\mathcal{P}^{jq}\partial_{p}\partial_{q}\ff \right)\,. \notag
\end{align}
The translation of the majority of the terms is straightforward,
$a^{(1)}$ is the first order perturbation of the scale factor taken as
$1/(1+z)$. To leading order, we can substitute $\Delta \ln a^{(1)} =
-\partial_{\chi}\left( v_{1i}n^{i}\right)$.

After several integrations by parts and translations between the
leading expression in our notation and the authors from
Ref.~\cite{cc1}, the main difference comes in the definition of the
convergence, where there is a numerical factor of
difference. Note that we are not taking into account the so called
`post-Born' contributions.
\begin{align}
&\Delta^{(2)}_{\text{Leading}} - {}^{(2)}\Delta^{\text{Bertacca}}_{\text{Leading}} \simeq  \notag\\
&\qquad\qquad \frac{\delta_{g}^{(1)}}{\chi}\int_{0}^{\chi_{s}}\dd\chi\left[ \nabla^{2}\left( \ff+\pf \right) + n^{i}n^{j}\left( \ff+\pf \right)_{,ij}+\frac{2}{\chi}\frac{\dd\dnuf}{\dd\varsigma}\right]  - 4\delta_{g}^{(1)}\kappa^{(1)}\notag\\
&\qquad\qquad +\frac{2}{\chi} \int_{0}^{\chi_{s}}\dd\chi\left[ \nabla^{2}\left( \ps+\fs \right) + n^{i}n^{j}\left( \fs+\ps \right)_{,ij}+\frac{2}{\chi}\frac{\dd\dnus}{\dd\varsigma}\right]- 2\kappa^{(2)}, \label{eq:dif-bertacca1} \\
&\approx 2 \delta_{g}^{(1)}\kappa^{(1)} - 4\delta_{g}^{(1)}\kappa^{(1)} + \kappa^{(2)} -2\kappa^{(2)}, \notag \\
& = - 2 \delta_{g}^{(1)}\kappa^{(1)}-\kappa^{(2)}.
\label{eq:dif-bertacca2}
\end{align}

\noindent In the second equality we used the definition given in Ref.~\cite{cc1} [see Eqs. (209) and (217) therein], to rewrite our notation into their convergence, and explicitly show the difference that arises from the convergence, where it appears to be adding twice as much to the galaxy overdensity.
%\textcolor{red}{does the above lead us to dismiss the bertacca expansion as the prefered expansion for analysing the second order?}
This last equation indicates that the difference with Ref. \cite{cc1} lies in factors of the first and second order lensing contributions, both of which represent relativistic effects. As we shall discuss below, the lensing terms may dominate the bispectrum signal at specific configurations, where the terms in Eq.~\eqref{eq:dif-bertacca2} must be considered carefully to avoid spurious contributions. 
%%%%%%%%%%%%%%%%%%%%%%%%%%%%%%%%%%%%%
\subsection{Comparison with Yoo and Zaldarriaga}
\label{subsec:yoozal2}
%%%%%%%%%%%%%%%%%%%%%%%%%%%%%%%%%%%%%

We now proceed to identify the terms of Ref.~\cite{yoo2}, which has to be put together from several different contributions since the full final result is not written down in closed form. The way the results are presented is one by one, so the main result -- the galaxy overdensity -- is not written explicitly; it is only shown in terms of previously found expressions, that the reader needs to find and then ``stitch-together'' in order to obtain the full expression. The main differences between our notation and Ref.~\cite{yoo2} are:
\begin{itemize}
\item Latin indices go from 0 to 3, Greek indices go from 1 to 3 (the opposite of our notation).
\item Metric perturbations are called $\mathcal{A}$ and $\mathcal{C}_{\alpha\beta}$, where $\mathcal{C}_{\alpha\beta} = \Psi \delta_{\alpha\beta}$ in our notation, for purely scalar perturbations. The second order perturbations are defined with no factor 1/2 at second order.
\item All perturbation orders are left implicit.
\item Work in a different basis that depends on angles, and leave some factors of $\left(\sin\theta\right)$ between expressions, which are not present in our derivation.
\end{itemize}
From \eq{eq:leadingdg}, we have that for Ref.~\cite{yoo2} the leading terms are
\be
{}^{(2)}\Delta_{\text{Leading}}^{\text{Yoo}} = \delta_{g}^{(2)} + \dVs + \delta_{g}^{(1)}\dVf.
\ee
The first order volume perturbation in Ref.~\cite{yoo2} is given by
\be
\dVf = -2 \kappa^{(1)} + H_{z}\partial_{z}\delta r^{(1)} = -2\kappa^{(1)} + \frac{1}{\H}\frac{\dd\left( v_{1i}n^{i} \right)}{\dd\varsigma},
\ee
where $\partial_{z} = H_{z}^{-1}\partial_{\chi}$. In Ref.~\cite{yoo2} the second order perturbation of the volume is
\be
\dVs = \delta \mathcal{D}_{L}^{(2)} + H_{z}\partial_{z}\delta r^{(2)} -2H_{z}\kappa^{(1)}\partial_{z} \delta r^{(1)} + \Delta x^{(1)b}\partial_{b}\dVf,
\ee
where $\delta\mathcal{D}_{L}^{(2)}$ is the luminosity distance, which is related to $\das$ through the \textit{Etherington's reciprocity theorem}, which is expressed as $d_{L} = \left( 1+z \right)^{2}d_{A}$. Lastly we need to expand 
\be
\Delta x^{(1)b}\partial_{b}\dVf = -2\nabla^{a}\ff\nabla_{a}\kappa^{(1)} + \frac{1}{\H}\nabla^{a}\ff\nabla_{a}\frac{\dd\left( v_{1i}n^{i} \right)}{\dd\varsigma} + \frac{1}{\H^{2}}\left(\frac{\dd\left( v_{1i}n^{i} \right)}{\dd\varsigma}\right)\left(\frac{\dd^{2}\left( v_{1i}n^{i} \right)}{\dd\varsigma^{2}}\right).
\ee
All these terms are already accounted for in $\delta\mathcal{D}_{L}^{(2)}$ and in $\delta r^{(2)}$. To avoid the cumbersome rewriting of the full term, we refrain from rewriting the terms here. For reference they are given in Eqs.~(78) and (50) in Ref.~\cite{yoo2}, respectively. Here, in agreement with Ref.~\cite{nielsen}, it can be seen that with these two terms, there is a double-counting effect, so this is the only difference between our leading terms, if we ignore this, then we are in agreement:
\begin{align}
\label{eq:diff-yoo}
\Delta^{(2)}_{\text{Leading}} - {}^{(2)}\Delta^{\text{Yoo}}_{\text{Leading}} &\simeq -2\nabla^{a}\ff\nabla_{a}\kappa^{(1)} + \frac{1}{\H}\nabla^{a}\ff\nabla_{a}\frac{\dd\left( v_{1i}n^{i} \right)}{\dd\varsigma} \\
&\qquad\qquad\qquad+ \frac{1}{\H^{2}}\left(\frac{\dd\left( v_{1i}n^{i} \right)}{\dd\varsigma}\right)\left(\frac{\dd^{2}\left( v_{1i}n^{i} \right)}{\dd\varsigma^{2}}\right). \notag
\end{align}
Thus we show explicitly that, while the leading terms are in complete agreement, the above terms are accounted for twice inside the luminosity distance and the radial perturbation.

Note that, in this case, our discrepancy with the approximation of Ref.~\cite{yoo2} lies in three terms, each with an independent origin and effect in observables. The first term in Eq.~\eqref{eq:diff-yoo} corresponds to the product of the first order lensing convergence times the transverse displacement. The second and third terms are contributions from the first order RSD combined with the transverse displacement, and a first order Taylor expansion of the RSD itself, correspondingly. 
%\textcolor{red}{Should we then dismiss this other expansion in favour of Di Dio or Ours?}
An important observable where the second order number counts are relevant is the matter bispectrum. The contribution from the dominant terms discussed in this section to such observable has been computed numerically in \cite{didio2015}. There, it is shown that for triangulations computed at a single redshift (or a narrow redshift bin), the result is dominated by standard Newtonian terms, that is the second order density and RSD terms of Eqs.~\eqref{eq:delta-2}, \eqref{eq:RSD-2} and their cross contribution in Eq.~\eqref{eq:delta-RSD}. However, for triangulations with vertices in more than one redshift, and redshifts separated by $\Delta z > 0.1$ the Newtonian terms rapidly decay and the bispectrum is dominated by the lensing contributions, mostly those of Eq.~\eqref{eq:k-2} (see also \cite{DiDio:2018unb}). In this sense the differences of Eq.~\eqref{eq:dif-bertacca2}, between our number counts estimation and those of Ref.~\cite{cc1}, would be manifest in the bispectrum calculated in triangles comprising more than one redshift bin. On the other hand, the RSD contributions to the difference with Ref.~\cite{yoo2} may result in spurious contributions to the bispectrum computed at a single redshift. A more detailed, numerical comparison including all contributions to the second order number counts is left for future work. 
%%%%%%%%%%%%%%%%%%%%%%%%%%%%%%%%%%%%%
\section{Discussion}
\label{sec:discussion}
%%%%%%%%%%%%%%%%%%%%%%%%%%%%%%%%%%%%%

In this paper we performed a comparison of the main result of
Ref.~\cite{FMH}, the galaxy number count at second order in
perturbation theory, with other derivations in the literature. The
approaches taken by different groups lead to slight differences beyond
linear order. We provide a comparison of the leading terms at second
order in Section \ref{sec:secondcomparison}, finding that our approach
and the previous works differ by numerical factors with one group, and
that another group double count some of the effects, leading to a
more significant difference with our expression. Once we take these
differences into account and correct for these typos, our results
are in agreement with the other groups (at leading order).
Although we do not provide a full comparison of the complete
expression, we will return to this issue in the future. Our result is
in agreement with the previous comparison made in Ref.~\cite{nielsen}
concentrated on the terms which are dominant in powers of $k/\H$,
i.e.~of the order $(k/\H)^4$.

Since the expressions found in the literature for the galaxy number
count agree with our result at leading order -- once the different notations and conventions have been taken into account and allowing for the typos discussed in Section \ref{sec:secondcomparison} -- all
of these results can be used for future and current surveys if the
leading order meets the required precision. On the other hand, beyond
leading order there are differences in the galaxy number count
results, which must be carefully considered when trying to estimate the
signal associated to primordial Non-Gaussianity encoded in the
powerspectrum and bispectrum (see e.g.~Ref.~\cite{Martinezetal}). To
what extent such differences affect the signal estimation is a task
best addressed numerically given the complexity of the expressions. This lies beyond the scope of the present paper and we shall return to
this point in the future.

In conclusion, our comparison shows that the differences between our
approach and previous studies presented in the literature lie mostly
on some of the geometrical definitions used on each calculation,
particularly with the direction of observation that could affect the
description of the peculiar velocities, which could bring extra (or
reduced) terms in the second order expansions, and result in the
addition of double terms of the velocities perturbations when
contrasted against our number counts. Once these differences have been
taken into account, the results agree at leading order as shown in
Section \ref{sec:secondcomparison}.
%This is particularly evident in the comparison with Ref.~\cite{yoo2}. 
While differences beyond leading order are not relevant yet, given the current
precision in galaxy catalogues, it is important to be prepared for
future surveys that will require accurate results beyond leading order. 
% \textcolor{red}{The referee wants us to advocate for a specific group? }

\acknowledgments

The authors are grateful to Pedro Carrilho, Chris Clarkson, and
Timothy Clifton for useful discussions and comments. JF acknowledges
support of studentship funded by Queen Mary University of London as
well as CONACYT grant No.~603085.  KAM is supported in part by the
STFC under grants ST/M001202/ and ST/P000592/1. JCH acknowledges
support from research grants SEP-CONACYT CB-2016-282569 and
PAPIIT-UNAM NI107521 \textit{Sector Oscuro y Agujers negros
  primordiales}. The tensor algebra package \texttt{xAct} \cite{xact}
and its subpackage \texttt{xPand} \cite{xpand} were employed to derive
the results presented.

%%%%%%%%%%%%%%%%%%%%%%%%%%%%%%%%%%%%%%%%%%%%%%%%%%%%%%
{
\bibliographystyle{ieeetr}

}

%%%%%%%%%%%%%%%%%%%%%%%%%%%%%%%%%%%%%%%%%%%%%%%%%%%%%%
\end{document}